\documentclass[twocolumn, aps, showpacs]{revtex4}

\input psfig.sty
\begin{document}

\title{Computationally efficient phase-field models with interface kinetics}
\author{Kalin Vetsigian and Nigel Goldenfeld}

\affiliation{Department of Physics, University of Illinois at
Urbana-Champaign
\\1110 West Green Street\\
Urbana, IL, 61801-3080}

\date{\today}

\begin{abstract}
We present a new phase-field model of solidification which allows
efficient computations in the regime when interface kinetic effects
dominate over capillary effects. The asymptotic analysis required to
relate the parameters in the phase-field with those of the original
sharp interface model is straightforward, and the resultant phase-field
model can be used for a wide range of material parameters.
\end{abstract}


\maketitle

\newcommand{\half}{\frac{1}{2}}
\newcommand{\dt}{\partial_t}
\newcommand{\dxi}{\paritial_{\xi}}
\newcommand{\beq}{\begin{equation}}
\newcommand{\eeq}{\end{equation}}

Phase-field model techniques have become increasingly recognized
as the tool of choice for solving moving free boundary
(sharp-interface) problems, and in particular solidification
processes \cite{Langer-first, Karma_RC, Goldenfeld, Wheeler}.
Following early work that related the phase-field models to the
original sharp interface model in the limit of zero interface
width \cite{Caginalp}, Karma and Rappel showed that calculations
could be performed when the interface width is of the order of the
capillary length \cite{Karma, Karma1}, even scaling with the
dendritic tip radius at low undercooling. This work provided a
particular relation between the parameters of the original sharp
interface model and the phase-field model, and has been the
starting point for accurate computations of solidification in the
important regime when capillary effects dominate interface
kinetics \cite{Casademunt}.

Recently however, there has been a growing interest in
the opposite regime where interface kinetics are dominant. This
interest is stimulated by experimental observation of the puzzling
morphological transition of the solidification front of Ni at high
undercooling \cite{Willnecker, Lum, Matson, Hofmeister}.

The purpose of this paper is to present a modification of the
phase-field model of solidification so as to enable efficient
computations in the regime when interface kinetics are the
dominant factor. The modification allows one to use an interface
thickness many times larger than the capillary length. The
methodology for performing the asymptotic analysis is different
and much easier to perform systematically to high order than the
asymptotic matching employed in all previous analyses, and is
capable of being used in other free boundary problems.

{\it Sharp interface model:-} The symmetric model for the
solidification of a pure melt from the liquid (L) phase to the solid
phase (S) is defined by the equations:
\begin{eqnarray}
\partial_t  u = D \nabla^2 u \\
\label{heat conservation}
\partial_n{u}|_S - \partial_n{u}|_L = {V}/{D} \\
\label{Gibbs-Thompson} u_i + d_0 k = -\mathcal{B}(V) \ .
\end{eqnarray}
Here $u=(T-T_M)c/L$ is the dimensionless temperature and $u_i$ is
its value at the solidification front, with $T$ being the
temperature in the liquid or solid, $T_M$ being the melting
temperature of a planar interface, $c$ being the specific heat and
$L$ being the latent heat of fusion per unit volume.  The
curvature of the solidification front is given by $k$ and the
capillary length is $d_0$.  $D$ is thermal diffusivity (assumed
here to be the same in both phases), and $V$ is the normal
velocity of the front. Equation (\ref{heat conservation})
expresses heat conservation at the interface, and equation
(\ref{Gibbs-Thompson}) is a modified Gibbs-Thomson condition,
which is a statement of local equilibrium at the interface with
the attachment kinetics included through the term
$\mathcal{B}(V)$. Traditionally, a linear kinetic undercooling
$\mathcal{B}(V)=\beta V$ is used. It should be stressed, however,
that the linearity of $u_i$ with respect to velocity is a purely
phenomenological assumption, and molecular dynamics simulations
\cite{Hoyt, Hoyt1} suggest that there are substantial deviations
from it at large undercooling. In this paper we are interested in
materials with large dimensionless parameter
$\tilde{\beta}\equiv{\beta D}/{d_0}$. This constant is a measure
of the importance of interface kinetics for a given material. It
takes very different values for different materials, and for Ni it
is estimated from molecular dynamics simulations to be as high as
90 \cite{Bragard}.

{\it The phase-field model:-} The phase-field equations can generally be
written in the form:
\begin{eqnarray}
\label{psi eq}
\tau\partial_t{\psi} &=& W^2\nabla^2\psi - f_{\psi}(\psi) - \lambda u g_{\psi}(\psi) \\
\partial_t{u} &=& D\nabla^2{u}+ \frac{1}{2}\partial_t h({\psi}).
\end{eqnarray}
Here $\psi$ represents the phase-field, $f(\psi)$ is a double well
potential, $g(\psi)$ shifts the relative height of the two minima
making one of the phases metastable for $u\neq0$.  Note that we have
used the subscript notation to denote differentiation: hence
$g_\psi(\psi)$ means $\partial g/\partial\psi$.  The sharp interface
boundary is recovered as the locus of points where $\psi=0$, and we are
interested in the behavior of the phase-field equations as the
phase-field interface width $W$ and relaxation time $\tau$ tend towards
$0$.  In order to solve the desired sharp interface model, we need to
ascertain what phase-field parameters $\{\tau, W, \lambda,
f(\psi),g(\psi)\}$ should be used given the values of $\{D, d_0, \beta$
or $ \mathcal{B}(V)\} $.

{\it Asymptotic analysis:-} The limit of small $W$ is a singular
one because $W$ multiplies a highest order derivative. All
previous works use asymptotic matching to deal with the
singularity. Here we will demonstrate a simpler approach based on
the fact that the equation for $u$ is linear so that it can be
trivially solved in terms of $\psi$. Following Karma and Rappel,
the analysis we give here focuses on the coordinate perpendicular
to the interface, $r$; the effect of the transverse dimensions is
incorporated by curvature dependent corrections, just as in their
work.

Requiring that $u$ is finite at $r\rightarrow\pm\infty$ and
requiring that $\psi(r)\rightarrow \mp 1$ sufficiently fast one
obtains
\begin{eqnarray}
\label{u_xi} u(\xi) &=& u_0 +
\frac{1}{2}pe^{-(p+q)\xi}\int_{-m\infty}^{\xi}{e^{(p+q)\eta}h(\psi(\eta))d\eta}\\
&=&u_0+\half p \left(\hat{u}+\frac{1}{p+q}\right)\ ,
\end{eqnarray}
where
\begin{equation}
q\equiv kW , \quad \ v\equiv\frac{V\tau}{W}\ll 1, \quad \
p\equiv\frac{V W}{D}, \quad \xi\equiv r/W,
\end{equation}
$m \equiv {\text sgn}(p+q)$, and the last line defines $\hat{u}$.
The dependence on $p+q$ expresses the singular nature of the
problem with respect to this parameter. From $u(\xi)$ one can
compute the outer limit $u_{out}(r)$ and thus obtain
$u_i=u_{out}(0) = u_0+ 1/2 \ {p}/(p+q)$.

Despite the singularity, if we are interested only in the profile
of $u$ near the interface one can expand in powers of $p$ and
$p+q$ and to first order obtain
\begin{equation}
\label{profile} u(\xi) = u_i +
\frac{1}{2}p\int_{-\infty}^\xi{\left( h(\psi(\eta))-1
\right)d\eta} + O\left(p(p+q)\right) \ .
\end{equation}
Using this expression and substituting it back in the equation for
$\psi$ we get an equation for $\psi$ which can be solved using regular
perturbation theory. In this way we recover the standard asymptotic
result of Karma and Rappel:
\begin{equation}
\label{Karma formula}
  W(\lambda) = \lambda \frac{d_o}{a_1}, \quad
 \ \tau(\lambda) = \lambda^2\left(\frac{\beta D}{d_0}+a_2
  \lambda\right)\frac{d_0^2}{a_1^2 D},
\end{equation}
where $a_1$ and $a_2$ are constants depending on $f(\psi)$ and
$g(\psi)$. Notice also that since $\psi$ is monotonic one can
compute the distance from the interface $\xi$ from $\psi$ and in
this way write to order $p$ \beq\label{u(xi)} u(\xi) = u_i + \half
p F_1\left(\psi;\{\psi(\xi)\}\right), \eeq where $F_1$ is defined
as the integral in equation (\ref{profile}).

From (\ref{Karma formula}) follows that with the functions
$f(\cdot)$, $g(\cdot)$, and $h(\cdot)$ fixed, $\lambda$ is the
only free parameter.

{\it Computational complexity:-} We now examine how the
computation time $t_c$ for the phase-field model scales with the
free parameter $\lambda$ in a discretized calculation with
adaptive mesh refinement and uniform grid elements in $d$
dimensions. Clearly, $t_c$ depends on the width of the phase-field
boundary layer $\tilde{W}$, the space resolution $\Delta{x}$, and
the time step $\Delta{t}$.  The inverse computation time scales
as: $(\Delta{t}/t)(\Delta{x}/{\tilde{W}})({\Delta{x}}/{L})^{d-1}$
where $L^{d-1}$ is the order of the surface area and $t$ is the
maximum time one wishes to evolve the system. For a spatially
explicit numerical scheme $\Delta{t} \leq \frac{1}{2}\tau
\left({\Delta{x}}/{W}\right)^2/{\lambda}$. The factor of $\lambda$
is included to guarantee accuracy in the presence of the term
$\lambda u g_\psi$. Collecting terms, and using $\tilde{W}\approx
W\propto\lambda d_0$ and $\tau\propto\lambda^2$ we obtain \beq
t_c^{-1}\propto\lambda^d\left(\frac{\Delta{x}}{W}\right)^d \eeq
showing that the computation time is highly sensitive to $\lambda$
and also depends on the spatial resolution required by the shape
of the interface profile of $\psi$: smoother profiles are better!

While it would be computationally efficient to work with large
$\lambda$, doing so would introduce higher order terms in the curvature
$k$ and velocity $V$ into the Gibbs-Thomson condition
(\ref{Gibbs-Thompson}). In principle, it might be possible to fine tune
$g(\psi)$, $f(\psi)$, and $h(\psi)$ to kill terms of order $p^2$, $pq$,
$v^2$, $vq$, etc. However, the resultant expressions are very
complicated, the integrals involved cannot be done analytically, there
are many terms to consider, and even if successful, this would most
likely introduce delicate fine structure into the phase-field profile
which would offset the computational benefits.

How can we do better?  Using equation (\ref{Karma formula}) we see that
we require
$
v = (\tilde{\beta} + a_2 \lambda) p \ll 1
$
which puts a very severe constraint on $p$ if $\tilde{\beta}$ is large.
Computationally this would be more important at large undercooling when
a thin temperature boundary layer forms around the solidification
front, and correspondingly $q<p$ so that the smallness of $p$ is the
limiting factor. (For example at undercooling $\Delta=0.8$ the theory
predicts ${q}/{p}={1}/{7}$ at a steady state dendrite tip.)
Therefore, a good objective is to modify the phase-field in a way that
relaxes the constraint on $v$.

A step in that direction was made by Bragard {\it et
al.} \cite{Bragard}, who replaced $\lambda u$ by
$H(\lambda u)$ in equation (\ref{psi eq}). $H(\cdot)$ is computed
numerically by solving the following non-linear eigenvalue
problem with appropriate boundary conditions on $\psi$:
\begin{equation}
\label{h of v eq.}
\frac{d^2\psi}{dx^2}-f_{\psi}(\psi)+v\frac{d\psi}{dx}-H(v)g_{\psi}(\psi)=0
\ .
\end{equation}
With $H(\cdot)$ chosen in this way the non-linearities appearing
in the standard phase-field model at large $v$ are cancelled by
the non-linearities in $H(\cdot)$. To relate the parameters they
use
\begin{equation}
\label{Bragard formula} d_0 = \frac{W}{\lambda}, \
\beta=\frac{\tau}{\lambda W}.
\end{equation}
However, this relationship is valid only in the limit of vanishing $p$,
i.e. when $u$ is approximately constant across the diffuse interface -
a result analogous to that of Caginalp for the standard phase-field.
Correspondingly, in \cite{Bragard} a value of $p$ close to $0.01$ is
used in computations.

Since $v+q$ is no longer a small parameter it is analytically more
involved to derive corrections for finite $p$. To compute the
linear part it is enough to consider small $v$ and the result is
\beq\label{brag+ formula} \tau = \lambda W(\beta +
a_2\frac{W}{D})\ \eeq which is the analog of Karma and Rappel's
formula.

Replacing (\ref{Bragard formula}) by (\ref{brag+ formula}) allows
much larger values of $p$ to be used and is a straightforward way
to make better use of the model proposed in \cite{Bragard}.
Numerically we observed that for $g_\psi=(1-\psi^2)^2$, the form
used in \cite{Bragard}, the non-linearities in $k$ and $V$ are
weak even for values of $v$ of the order of 20. However, the
Bragard {\it et al.} phase-field model, even with the improved
asymptotics (\ref{brag+ formula}) that we derived, still does not
provide the desired degree of computational improvement because
the phase-field profile develops a new length scale of order
${W}/{H(v)}$ which needs to be resolved numerically in order to
avoid artifacts. In addition, $H$ increases very rapidly with $v$.

{\it New class of models:-} We propose to replace $\tau$ by
$\tau_R(\psi)$ in such a way so that the effective equation for
$\psi$ becomes
\begin{equation}
\label{u_i psi} \tau\partial_t{\psi} = W^2\nabla^2{\psi} -
f_{\psi}(\psi) - \lambda u_i W|\nabla\psi| \ .
\end{equation}
What are the advantages of doing this? The asymptotic analysis is
greatly simplified because the equation for $\psi$ can be analyzed
separately from that for $u$. The solution for $\psi$ is simply
$\psi(\xi)= \psi_0(\xi)$ where $\psi_0(\xi)$ is the solution of
$\partial_{\xi}^2\psi_0-f_\psi(\psi_0)=0$, and the relation
between the parameters is simply given by (\ref{Bragard formula}).

Now we can rewrite (\ref{u_i psi}) as
\begin{equation}
\tau\partial_t{\psi} = W^2\nabla^2{\psi} - f_{\psi} - \lambda u
W|\nabla{\psi}| - \lambda \frac{p}{2} F_1(\psi) W|\nabla{\psi}|
\end{equation}
with the only problem being the presence of $p$ in the evolution equation.
The final trick is to express $p$ in terms of $\partial_t \psi$:
\begin{equation}
p = \frac{W}{D}V = -\frac{W}{D}\frac{\partial_t \psi}{\partial_{x}\psi} =
\frac{W}{D}\frac{\partial_t \psi}{|\nabla{\psi}|}.
\end{equation}
The equation for $\psi$ becomes \beq\label{our model}
\tau_R(\psi)\partial_t{\psi} = W^2\nabla^2{\psi} - f_{\psi}(\psi)
- \lambda u W|\nabla{\psi}|, \eeq where
\begin{equation}
\tau_R = \tau -  \frac{1}{2}\lambda \frac{W^2}{D} F_1(\psi) \ .
\end{equation}
For $f(\psi)=\frac{1}{4}\left(1-\psi^2\right)^2$ and
$h(\psi)=\psi$ we have $F_1(\psi) =
\sqrt{2}\ln\left((\psi+1)/{2}\right)$. It follows that
$\tau_R\geq\tau$ which means that the model is well behaved. The
expression for $\tau_R$ can be compared with Karma and Rappel's
formula which can be rewritten in the form $ \tau' = \tau + a_2
\lambda\frac{W^2}{D}$. $a_2$ is approximately the value of $-1/2
F_1(0)$. As it stands $\tau_R(-1)=\infty$ and points with
$\psi=-1$ cannot evolve, so some cutoff near $\psi=-1$ should be
introduced. Experiments show that results are insensitive to the
exact form of the cutoff.

The restriction of order $p$ accuracy comes from expanding
$u(\xi)$ near the interface. It is possible to go to higher orders
in $p$ or simply use the full expression (\ref{u_xi}). This would
result in an {\it implicit\/} equation for $\partial_t\psi$
\beq\label{implicit dpsi_dt} \tau\partial_t{\psi} =
W^2\nabla^2{\psi} - f_{\psi} - \lambda u W|\nabla{\psi}| -
\frac{\lambda W^2}{2D}\dt\psi\hat{u}(\psi,p+q) . \eeq The function
$\hat{u}$ can be tabulated in advance and equation (\ref{implicit
dpsi_dt}) can be solved iteratively at each time step.

Different approximations to eqn. (\ref{implicit dpsi_dt}) lead to
different schemes. For example if we consider the next order term in
(\ref{profile}) $\half p(p+q)F_2\left(\psi;\{\psi(\xi)\}\right)$ we end
up with a quadratic equation for $\dt\psi$. The model including $p^2$
corrections is:
 \begin{eqnarray}
 &\tau_R = \tau
- \frac{\lambda W^2}{2
D}\left(F_1(\psi)+qF_2(\psi)\right) \nonumber \\
&\tau_R(\partial_t\psi)_0=W^2\nabla^2{\psi} - f_{\psi}(\psi) -
\lambda W u \left|\nabla\psi\right| \nonumber\\
 &\alpha =
(\partial_t\psi)_0\frac{1}{\tau_R}\frac{\lambda
W^4}{2D^2}\frac{F_2(\psi)}{g_{\psi}(\psi)} \nonumber\\
\label{R grad p2}
\partial_t{\psi}&=(\partial_t\psi)_0\frac{1-\sqrt{1-4\alpha}}{2\alpha}
=(\partial_t\psi)_0(1+\alpha+2\alpha^2+\ldots).
\end{eqnarray}
In the evolution equation $q=Wk=W\nabla\cdot{\bf n}$ with ${\bf n}
= {\nabla{\psi}}/{|\nabla{\psi}|}$.

Another application of the technique is to correct for terms of order
$pq$ at small undercooling when $p\ll q$. In this regime terms of order
$pq$ are not negligible compared to terms of order $p$. To achieve this
it is enough to use $(\dt\psi)_0$ from above without correcting it.

The phase-field model can be generalized to handle arbitrary
interface kinetics $u_i = -d_0 k - \mathcal{B}(V)$ at order $p$
and arbitrary $v$. The resultant phase-field model equation is
\begin{eqnarray}
\tau_R(\psi, u)\partial_t\psi&=&W^2\left(\nabla^2{\psi}
-k(\nabla{\psi}\cdot{\bf n})\right)- f_{\psi}(\psi) \nonumber \\
\label{non lin phase}&-& \mathcal{B}^{-1}\left(\lambda(u+d_0 k)
\right) W\left|\nabla{\psi}\right| \\
\label{nl tau}\tau_R(\psi,u)&=&\tau-\frac{\lambda W^2}{2
D}F_1(\psi){\mathcal{B}^{-1}}'\left(\lambda \left(u+d_0
k\right)\right).
\end{eqnarray}

The above recipe for improving phase-field models can be used also
in cases when the $\psi$ profile changes with $V$ and $k$. For
example it can be applied to the model of Bragard {\it et al.} by
effectively replacing $u$ with $u_i$ yielding \beq\label{R
g5}\tau_R(\psi, u)\partial_t{\psi} = W^2\nabla^2{\psi} -
f_{\psi}(\psi) - H(-\lambda u) g_{\psi}(\psi),\eeq with \beq\tau_R
= \tau + \frac{\lambda W}{2D}H'(-\lambda
u)\tilde{F_1}\left(\psi,H(-\lambda
u)\right)\frac{g_{\psi}(\psi)}{|\nabla{\psi}|} \ . \eeq In this
case the expression for $\tau_R$ is more complicated because
$\psi$ changes with $v$. The functions $H(v)$ and $\psi_v(\xi)$
which solve equation (\ref{h of v eq.}) and $F_1$ can be
pre-computed numerically leading to a very efficient numerical
scheme.

{\it Anisotropy:-} To include anisotropy we need only to replace
$W$ with $W({\bf n})$, $\tau$ with $\tau({\bf n})$, and in three
dimensions, $W^2\nabla^2\psi$ with
$-\frac{\delta}{\delta\psi}\int{dV W({\bf n})^2(\nabla{\psi})^2}$
to obtain the Gibbs-Thomson condition
\begin{equation}
u_i = -\frac{1}{\lambda}\sum_{i=1,2}{\left[W({\bf
n})+\partial_{\theta_i}^2W({\bf n})\right]\frac{1}{R_i}} -
\frac{\tau({\bf n})}{\lambda W({\bf n})}V,
\end{equation}
where $\theta_1$ and $\theta_2$ are the angles between the normal
and the local principle directions on the interface, and
$R_1^{-1}$ and $R_2^{-1}$ $-$ the principle curvatures.

{\it Numerical Experiments:-} We now compare the performance of
different phase-field models in one-dimensional simulations. The benchmark
problem solved is $u(t=0,x)=-\Delta$ for $x\in(-\infty,\infty)$
with the solid-liquid interface initially at $x=0$. The interface
velocity, $V(t)$, is compared to that for the sharp interface
model obtained via direct numerical integration.

The models compared are identified as follows. $ST$: standard
phase-field model (\ref{psi eq}); $BR$: Bragard {\em et al.} model
with asymptotic relation (\ref{Bragard formula}); $BR+$: the above
model with the improved asymptotic relation (\ref{brag+ formula});
$\tau_R$: the new model (\ref{our model}); $\tau_R+p^2$: the new
model with $p^2$ corrections (\ref{R grad p2}); $\tau_R\ BR$: the
improved version of $BR$ given in (\ref{R g5}). We used
$h(\psi)=\psi$ and $g_{\psi}=(1-\psi^2)^2$.

As an example, we computed the velocity of the front after
$t=3.5\times 10^4 d_0^2/D$ for $\tilde{\beta}=10$, $\Delta = 1.2$
and $\lambda = 15$. The exact result is $V d_0/D=0.021$. The
results for models $ST$, $BR$, $\tau_R\ BR$ were 0.012, 0.044,
0.020 respectively, showing that the previously existing models
are inadequate in this regime.

\begin{figure}[!t]
\leavevmode\centering\psfig{file=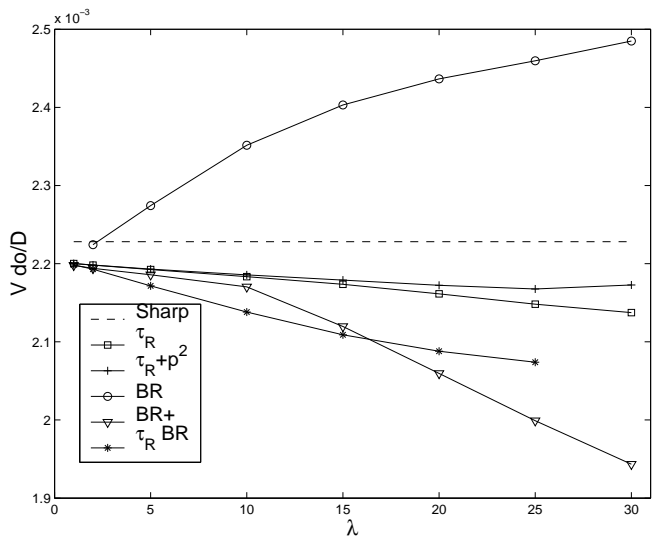,width=\columnwidth}
\caption{Interface velocity from different phase-field models as a
function of $\lambda$. $\tilde{\beta}=90$, $\Delta=1.2$. At
$t=8\times 10^6 d_0^2/D$. $\Delta{x}=0.5W=0.5\lambda d_0$}.
\label{V lambda}
\end{figure}

Figure~\ref{V lambda} compares the systematic deviations of
various phase-field models from the sharp interface solution as a
function of $\lambda$. One clearly sees that $BR$ model leads to
errors linear in $\lambda$. For $BR+$ unintended nonlinearities in
the Gibbs-Thomson condition quickly increase the error with
$\lambda$. In contrast $\tau_R$ and $\tau_R+p^2$ models yield
approximately the same velocity for the entire range of $\lambda$
considered. Using the values for Ni cited in \cite{Bragard},
$D=10^{-5}{\rm m^2/sec}$ and $d_0=5.56\times10^{-10}$, the choice
$\Delta=1.2$ corresponds to a steady state velocity
$V=(\Delta-1)D/(\tilde{\beta}d_0)=40$m/sec which is approximately
where the experimentally observed morphological transition occurs.

To match the accuracy of $\tau_R$ model with $\lambda=30$ we need to
take about $\lambda=5$ in $BR$ (measuring deviations from the limiting
phase field value). In 3D, this leads to about $(30/5)^3\approx 200$
times increase in computational speed as compared to the simulations in
\cite{Bragard}. If we use the Bragard {\it et al.} model with our improved
asymptotics, the new $\tau_R(\psi)$ models will be
about $3^3=27$ times faster. The above figures are just for
illustration, the precise computational gains will depend on the
desired accuracy and the regime of interest.

In conclusion, the models described here are the first that can
systematically handle interface kinetics dominated growth in and beyond
the thin-interface limit enabling huge gains in computational
efficiency.

We thank Jon Dantzig for useful discussions and his interest in this
work.  This work was supported in part by the National Science
Foundation through grants NSF-DMR-99-70690 and NSF-DMR-01-21695.

\bibliography{dendrites}

\begin{thebibliography}{10}

\bibitem{Langer-first}
J.S. Langer.
\newblock In G.~Grinstein and G.~Mazenko, editors, {\em Directions in condensed
  matter physics}, page 165. World Scientific Press, 1986.

\bibitem{Karma_RC}
A.~Karma and W.~Rappel.
\newblock {\em Phys. Rev. E}, 53:R3017, 1996.

\bibitem{Goldenfeld}
N.~Provatas, N.~Goldenfeld, and J.~Dantzig.
\newblock {\em Phys. Rev. Lett.}, 80:3308, 1998.

\bibitem{Wheeler}
W.~Wheeler, G.~McFadden, and W.~Boettinger.
\newblock {\em Proc. R. Soc. London A}, 452:495, 1996.

\bibitem{Caginalp}
G.~Caginalp and X.~Chen.
\newblock In M.E.Gurtin and G.B. McFadden, editors, {\em On the evolution of
  phase boundaries}, volume~1, page~1. Springer-Verlag, 1992.

\bibitem{Karma}
A.Karma and W.~Rappel.
\newblock {\em Phys. Rev. Lett.}, 77:4050, 1996.

\bibitem{Karma1}
A.~Karma and W.~Rappel.
\newblock {\em Phys. Rev. E}, 57:4323, 1997.

\bibitem{Casademunt}
For a~recent review see R. Gonz\'alez-Cinca~et al.
\newblock {\em cond-mat/0305058}.

\bibitem{Willnecker}
R.~Willnecker.
\newblock {\em Phys. Rev. Lett.}, 62:2707, 1989.

\bibitem{Lum}
J.~Lum, D.~Matson, and M.~Flemings.
\newblock {\em Metall. Mater. Trans. B}, 27:865, 1996.

\bibitem{Matson}
D.~M. Matson.
\newblock In S.P.Marsh, J.A. Dantzig, R.~Trivedi, W.~Hofmeister, M.G. Chu, E.J.
  Lavernia, and J.~H. Chun, editors, {\em Solidification 1998}, volume~1, page
  233. The Mineral, Metals and Materials Society, 1998.

\bibitem{Hofmeister}
W.~Hofmeister, R.~Bayuzick, and M.~Robinson.
\newblock {\em Rev. Sci. Instrum.}, 61:2220, 1990.

\bibitem{Hoyt}
J.~Hoyt, B.~Sadigh, M.~Asta, and S.~Foiles.
\newblock {\em Acta mater.}, 47:3181, 1999.

\bibitem{Hoyt1}
J.~Hoyt, M.~Asta, and A.~Karma.
\newblock {\em Phys. Rev. Lett.}, 86:5530, 2001.

\bibitem{Bragard}
J.~Bragard, A.~Karma, Y.~Lee, and M.~Plapp.
\newblock {\em Interface Science}, 10:121, 2002.

\end{thebibliography}
\bibliographystyle{unsrt}

\end{document}